\documentclass[twocolumn,showpacs,superscriptaddress,prl]{revtex4}
\usepackage{graphicx}% Include figure files
\usepackage{dcolumn}% Align table columns on decimal point
\usepackage{bm}% bold math
\usepackage{amsmath,amssymb}

\def\be{\begin{equation}}
\def\ee{\end{equation}}
\def\e#1{\label{#1}\end{equation}}
\def\bea{\begin{eqnarray}}
\def\eea{\end{eqnarray}}
\def\ea#1{\label{#1}\end{eqnarray}}

\def\bem#1{\begin{mathletters}\label{#1}}
\def\eml{\end{mathletters}}

\def\ket#1{{|#1\rangle}}

\def\4#1{{\boldsymbol{#1}}}
\def\8#1{{\widetilde{#1}}}

\def\braket#1#2{{\langle#1|#2\rangle}}
\def\mean#1{{\langle#1\rangle}}
\begin{document}

%\preprint{APS/123-QED}

\title{GENERALIZED TELEPORTATION PROTOCOL}

\author{Goren Gordon}
\affiliation{Department of Chemical Physics,
Weizmann Institute of Science, Rehovot 76100, Israel}

\author{Gustavo Rigolin}
\email{rigolin@ifi.unicamp.br}
\affiliation{Departamento de F\'{\i}sica da Mat\'eria Condensada,
Instituto de F\'{\i}sica Gleb Wataghin, Universidade Estadual de
Campinas, C.P.6165, cep 13084-971, Campinas, S\~ao Paulo, Brazil}

\date{\today}

\begin{abstract}
A generalized teleportation protocol (GTP) for $N$ qubits is
presented, where the teleportation channels are non-maximally
entangled and all the free parameters of the protocol are
considered: Alice's measurement basis, her sets of
acceptable results, and Bob's unitary operations.
The full range of Fidelity ($F$) of the teleported state and the
Probability of Success ($P_{suc}$) to obtain a given fidelity are achieved by
changing these free parameters. A channel efficiency bound is found,
where one can determine how to divide it between $F$ and
$P_{suc}$. A one qubit formulation is presented
and then expanded to $N$ qubits. A proposed experimental setup
that implements the GTP is given using linear optics.
\end{abstract}
\pacs{03.67.Mn, 03.67.Hk, 03.67.-a}

\keywords{Quantum teleportation; entangled channels; measurement
basis }

\maketitle
%%
%%\section{Introduction}

The concept of entanglement is central in quantum information
processing. One major breakthrough was obtained by Bennett
\textit{et al.} \cite{ben93}, who created the quantum
teleportation protocol. Right after its proposal, Bouwmeester
\textit{et al.} \cite{bou97} and Boschi \textit{et al.}
\cite{bos98} experimentally implemented the teleportation
protocol. Interesting extensions were subsequently proposed,
specially those regarding the teleportation of more than one qubit
\cite{ikr00}. All the previous proposals assume, nevertheless,
that the quantum channels used to teleport the qubits are
noiseless maximally entangled states. But in a realistic scenario
noisy effects and decoherence decrease the entanglement of the
channel. In this scenario, Agrawal and Pati \cite{aga02}
constructed a protocol where it is possible to achieve unity
fidelity teleportation of one qubit using directly non-maximally
entangled channels. The price to pay is that the protocol is no
more deterministic.

This paper generalizes Agrawal and Pati \cite{aga02} work and
expand it to teleport N qubits using directly N non-maximally
entangled channels. The two previous proposals, namely the
standard protocol \cite{ben93} and the probabilistic protocol
\cite{aga02} are generalized and written in one single formalism.
In this generalization one can enhance $P_{suc}$,
on expense of $F$, by using a different measurement
basis. The total channel efficiency is bounded by the entanglement
of the two qubit channels, but one can decide how to ``divide''
this bound between $F$ and $P_{suc}$ to
obtain this fidelity, according to the system requirements.

In general, Alice may wish to teleport $N$ qubits. A $N$ qubit
state has $2^{N}$ arbitrary unknown complex amplitudes. Let
$\alpha_i$, where $i=1, ..., 2^N$, represent these amplitudes.
Alice has one channel per qubit to be teleported. We assume that
each channel is composed of two entangled qubits, one with Alice
and one with Bob. The channels need not be maximally entangled and
their entanglement are parametrized by $n_k$, where $k=1,...,N$.
The protocol involves measurements by Alice, meaning that she uses
a specific measurement basis to project her $2N$ qubits ($N$
qubits she wishes to teleport plus $N$ qubits from the $N$ two
qubit channels). The measurement basis is characterized by the
parameters $m_s$, where $s=1,...,N$. The measurement yields
different possible results $\ket{R_j}$, each with probability
$P_j$, where $j=1, ..., 2^{2N}$, due to the fact that the
measurement is performed jointly on the qubits to be teleported
and the channel qubits. Alice decides, beforehand, which results
are acceptable, i.e. the protocol has succeeded, and which results
are not, meaning the protocol has failed. The acceptable results,
$\ket{R_l}$, where $\{l\}\subseteq\{j\}$, she transmits to Bob via
a classical channel. In terms of the measurement basis, the total
initial state can be written as
\be
\ket{\psi}=\underbrace{\sum_l\beta_l\ket{R^A_l}\ket{\psi^B_l}}_{\textrm{Success}}+
\underbrace{\sum_{j\neq l}
\beta_j\ket{R^A_j}\ket{\psi^B_j}}_{\textrm{Failure}},
\label{Alice-Wavefunction}
\ee
where $P_j=|\beta_j|^2$. In this scenario, Alice has
$P_{suc}=\sum_l P_l$ probability of success, meaning one of the
acceptable results has been obtained. After measurement, the
initial state collapses to one of the $\ket{R^A_j}\ket{\psi^B}$
states and Alice transmits to Bob her outcome, i.e. the value of
$j$, conditioned on the restriction $j \in \{l\}$. Bob now
performs a unitary transformation ${\bf U}_l$ on his $N$ qubits,
which can be different for each one of Alice's measurement
results. We assume Bob's unitary operations are local in his
qubits: $\textbf{U}_{l}=U_{1} \otimes \cdots \otimes U_{N}$. A
general unitary transformation on $N$ qubits can be represented by
$4N$ parameters (four parameters for each local unitary operation)
and Bob must decide beforehand what operations to do on his qubits
conditioned on the information received from Alice.  However, for
each result Bob receives from Alice, he can choose among the $4N$
parameters. (These parameters are part of the protocol and cannot
be changed during the teleportation.) After these transformations,
Bob obtains the final state $\ket{\phi^B_l}$, with the
accompanying fidelity $F_l=|\braket{\phi^A}{\phi^B_l}|^2$.

The quantities of interest here, i.e. probability and fidelity,
are dependent on $|\alpha_i|^2$ and $|\alpha_i\alpha_{j}|^2$.
However, we wish to get $\alpha_{i}$-independent results for the
protocol. Since the input state is arbitrary and in general
unknown, this is achieved by averaging over these quantities with
the appropriate distribution function. This is done by using
spherical coordinates in a $2^{N}$-dimensional real space, where
$N$ is the number of qubits to be teleported. We thus find that
$\langle|\alpha_{i}|^{2}\rangle = 1/2^{N}$ and $\langle
|\alpha_{i}\alpha_{j}|^2\rangle=(2^{\delta_{ij}-\;N})/(1+2^{N})$,
which is all that is required in the following calculations. Here,
$\delta_{ij}=1$ if $i=j$ and zero otherwise.

Alice's probabilities $P_j$ may depend on the particular state to
be teleported ($\alpha_{i}$), thus we use the average probability
$\mean{P_j}$. Bob gets the fidelities $F_l$ with probability
$P_l$. Averaging over many implementations of the protocol we
obtain the protocol efficiency $C^{pro}$, which can also be viewed
from a different perspective by defining the protocol fidelity
$F^{pro}$. The channel efficiency, $C^{channel}$, is defined as the
maximal protocol efficiency, where the maximization is done over all
the free parameters (which exclude $\{n_k\}$).
\bea
&&C^{pro}=\sum_l \mean{P_lF_l}. \label{defCpro}\\
&&F^{pro}=\frac{C^{pro}}{\mean{P_{suc}}}= \frac{\mean{\sum_l
P_lF_l}}{\mean{\sum_l P_l}}. \label{Fpro}\\
&&C^{channel}(\{n_k\})=\max_{m_s,\;\{l\},\;\mathbf{U}_{l}}C^{pro}.
\eea
The protocol efficiency can be interpreted as the average qubit
transmission rate for a specific protocol choice and is the
product of the probability of its success and its fidelity. For
the specific case where Alice accepts all results: $P_{suc}=1$ and
$C^{pro}=F^{pro}$. Eq.~(\ref{Fpro}) shows that Alice and Bob have
the freedom to modify $F$ and $P_{suc}$
while maintaining the same protocol efficiency. For a given
$C^{pro}$, they can get higher (lower) fidelity lowering
(increasing) $P_{suc}$. The channel efficiency
gives the maximal qubit teleportation rate for a given channel.

For the one qubit case, a quantum channel which is not maximally
entangled (we consider pure states only) 
is given as \cite{aga02} $\ket{\Phi_{n}^{+}} =
(1/\sqrt{1+|n|^2})(\ket{00} + n \ket{11})$.  Here $n$ is a
co\-m\-plex nu\-m\-ber in whi\-ch $0 \leq |n| \leq 1$. The concurrence
 for this state, a well known entanglement monotone \cite{woo98},
is $c(n) = 2|n|/(1+|n|^{2}$), which is a monotonically increasing
function of $|n|$. (Throughout the paper when we talk about the 
degree of entanglement of a state we are referring to its concurrence.)
Alice wishes to teleport the qubit  $\ket{\phi^A}=\alpha_1
\ket{0}+\alpha_2\ket{1}$, where $|\alpha_1|^{2}+|\alpha_2|^{2}=1$.
Alice's  arbitrary Bell-measurement basis are
$\{\ket{R_j}\}=\{\ket{\Phi_{m}^{+}},\ket{\Phi_{m}^{-}},\ket{\Psi_{m}^{+}},
\ket{\Psi_{m}^{-}}\}$, where
$\ket{\Phi_{m}^{+}}=M(\ket{00}+m\ket{11})$,
$\ket{\Phi_{m}^{-}}=M(m^{*}\ket{00}-\ket{11})$,
$\ket{\Psi_{m}^{+}}=M(\ket{01}+m\ket{10})$, and
$\ket{\Psi_{m}^{-}}=M(m^{*}\ket{01}-\ket{10})$. Here
$M=1/\sqrt{1+|m|^2}$ and $m^{*}$ is the complex conjugate
of $m$.

The initial three-qubit state (Alice's qubit and the channel
qubits) can be projected onto Alice's two qubits arbitrary
Bell-basis, with the appropriate probabilities. Alice transmits
the result of her measurement via a classical channel to Bob, who,
whereupon, performs a unitary transformation on his qubit. Bob has
$16$ free parameters, four for each of Alice's measurement result.
We restrict ourselves, however, to only one free parameter
($\theta_j$) for each result. The unitary operations are
$\{\ket{R_j}\}\rightarrow\exp(i\sigma_z\theta_j)O_j$, where
$\{O_j\}=\{I,\sigma_z,\sigma_x,\sigma_z\sigma_x\}$. $I$ is the
identity and $\sigma$ are the usual Pauli matrices.

Implementing the averaging procedure des\-cri\-bed a\-bo\-ve, the averaged
probabilities and fidelities are found to be
\bea
\mean{P_{\Phi^+_m}} = \mean{P_{\Psi^-_m}} &=&
\frac{1+|nm|^{2}}{2(1+|n|^2)(1+|m|^2)}, \\
\mean{P_{\Phi^-_m}} = \mean{P_{\Psi^+_m}}
&=&\frac{|n|^{2}+|m|^{2}}{2(1+|n|^2)(1+|m|^2)}. \\
\mean{F_{\Phi^+_m\!,\Psi^-_m}P_{\Phi^+_m\!,\Psi^-_m}}
\!\!\!&=&\!\!\!
\frac{1\!+\!|nm|^2\!+\!|mn|\!\cos(\xi_{\Phi^{+}\!,\Psi^{-}})}{3(1+|n|^2)(1+|m|^2)},
 \label{fide1}\\
\mean{F_{\Phi^-_m\!,\Psi^+_m}P_{\Phi^-_m\!,\Psi^+_m}}
\!\!\!&=&\!\!\!
\frac{|n|^2\!+\!|m|^2\!+\!|mn|\!\cos(\xi_{\Phi^{-}\!,\Psi^{+}})}{3(1+|n|^2)(1+|m|^2)},
\label{fide2} \eea
where, using that $n=|n|\mathrm{e}^{\mathrm{i}\theta_{n}}$ and
$m=|m|\mathrm{e}^{\mathrm{i}\theta_{m}}$, we have
$\xi_{\Phi^{\pm}} = \theta_{n} - \theta_{m} -
2\theta_{\Phi^{\pm}}$ and $\xi_{\Psi^{\pm}}
= \theta_{n} + \theta_{m} + 2\theta_{\Psi^{\pm}}$.%

For the special case where Alice accepts all possible results,
i.e. the protocol always succeed, $P_{suc}=1$ and the protocol
efficiency is:
\be
C^{pro}=\frac{2}{3}\left(1+\frac{|nm|\sum_{j=\Phi^\pm,\Psi^\pm}\cos(\xi_j)}
{2(1+|n|^2)(1+|m|^2)}\right). \label{Cpro} \ee

Looking at Eq.~(\ref{Cpro}) we see that $C^{pro}$ is maximum if
$\xi_{j}=2\pi q_{j}$, $q_{j}$ integer. This can always be achieved
if Bob properly adjusts his four free parameters $\theta_{j}$,
which depend on the channel and measuring basis entanglement
($n_{k}$ and $m_{s}$, respectively, assumed to be known by Bob).
This is equivalent to working with real $n$ and $m$, a scenario
which, unless stated otherwise, is assumed throughout the rest of
this paper. Therefore, Eq.~(\ref{Cpro}) reads
$C^{pro}=(2/3)(1+ c(n)c(m)/2)$, where $c$
is the concurrence. Note that Eq.~(\ref{Cpro}) is invariant if we
interchange the parameters $m$ and $n$. This remarkable result
shows that $C^{pro}$ {\em is the same if we exchange the channel
entanglement and the measuring basis entanglement}. We can easily
see that the channel efficiency, i.e. maximal protocol efficiency when
$m$ and $\xi_j$ are the free parameters, is achieved for all $n$
at $m=1$ and all $\xi_j=0$.

We can consider the case of a dephased channel, where the quantum state
describing it accumulates a relative phase. In the
GTP notation, this amounts to $n=e^{i\theta_n}$. Assuming the
dephasing rate is known, this obstacle can be overcome by an
appropriate unitary operation performed by Bob. Let us assume, for
example, $m=1$. We see that by performing unitary transformations
such that $\theta_{\Phi^\pm}=\theta_n/2$ and
$\theta_{\Psi^\pm}=-\theta_n/2$, we eliminate the dephasing and it
results in a unity fidelity teleportation protocol (no averaging
required). This result shows that {\em only the entanglement of
the channel} is important for the teleportation protocol to
succeed and {\em not which entangled state is used}.

In the standard protocol \cite{ben93}, Alice uses the standard
Bell-basis (maximally entangled states) to implement her joint
measurements. In the GTP formulation, this corresponds to $m=1$
and all $\xi_j=0$. This results in $P_{suc}=1$, $C^{std}=F^{std}=
(2/3)(1+n/(1+n^2))$. In the probabilistic
quantum teleportation (PQT) protocol \cite{aga02} Alice uses a
special measurement basis, which in the GTP formalism corresponds
to real $m=n$ and all $\xi_j=0$. Also
$\{\ket{R_l}\}=\{\ket{\Phi^-_n},\ket{\Psi^+_n}\}$ which results in
$P_{suc}=2n^2/(1+n^2)^2$, $F^{PQT}=1$, and
$C^{PQT}=2n^2/(1+n^2)^2$.

\begin{figure}[!htb]
\centering\includegraphics[width=8.0cm]{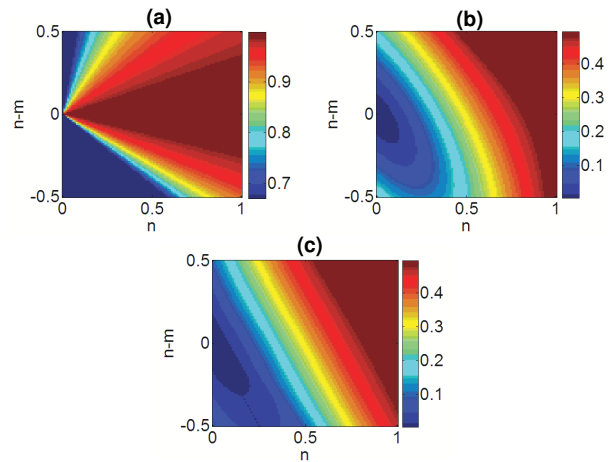}
 \caption{(Color online) PQT attributes as a function
of the dimensionless parameter $n$ and the perturbation from UFP,
i.e. $n-m=0$. (a) $F^{PQT}$; (b) $P_{suc}$; (c) $C^{PQT}$.}
 \protect\label{Fig-PQT}
\end{figure}

As seen from these examples, we can create a tradeoff between the
fidelity of the protocol and the probability of its success. (We
assume all $\xi_j=0$). When Alice decides not to accept all
possible results, i.e. not to transmit all the results to Bob, the
protocol will have less than unity $P_{suc}$. However, as shown in
the probabilistic quantum protocol, we gain unit fidelity when the
teleportation does succeed under a special circumstance ($m=n$).
It is noteworthy to consider the perturbed case of this protocol,
i.e. $m\simeq n$. This requires averaging and results in less than
unit fidelity. Fig.~\ref{Fig-PQT} show the perturbation in
protocol fidelity $F^{PQT}$ (Eq.~(\ref{Fpro})), probability of
success $P_{suc}$, and the protocol efficiency $C^{PQT}$
(Eq.~(\ref{defCpro})) as a function of $n$ and the perturbation
from the Unity Fidelity Protocol (UFP), i.e. $n-m=0$. As can be
seen, we lose fidelity as the perturbation grows (Fig.
\ref{Fig-PQT}(a)), but $P_{suc}$ is enhanced (Fig.
\ref{Fig-PQT}(b)). The mean fidelity grows as $m \rightarrow 1$,
as in the general case. We should note that this scenario is more
realistic since the entanglement in the channel is not known
completely, implying that the measurement basis cannot be set to
$m=n$, but only as a close approximation.

The generalized teleportation protocol detailed above will now be
expanded to $N$ qubits. (It can be seen as $N$
single qubit protocols implemented at once or in sequence. However, the 
overall fidelity and protocol efficiency are not trivial extensions of 
previous results.) 
The state Alice wants to teleport is the
most general pure state for N qubits: $|\phi^{A}\rangle =
\sum_{i=1}^{2^{N}}\alpha_{i}|\text{Bin}(i-1)\rangle$,
where $\text{Bin}(i)$ is the binary representation of the integer
$i$ with zeros padded to its left in order to leave all binary
numbers with the same amount of digits. Now Alice needs N
two-qubit channels, which is given by N Bell states with different
degrees of entanglement (in general $n_{i}\neq n_{j}$, for $i\neq
j$): $|\phi^{\text{channel}}\rangle =
\bigotimes_{i=1}^{N}|\phi_{n_{i}}^{+}\rangle$. For each Bell
state, one qubit is with Alice and the other one with Bob.

The rest of the protocol is similar to the one-qubit protocol: (a)
Alice performs $N$ Bell measurements. The states expanding each
basis she projects need not have the same degree of entanglement
($m_{i} \neq m_{j}$ in general); (b) Alice informs Bob of the
acceptable results. At most she transmits $2N$ bits of classical
information to Bob, two bits for each Bell measurement considered
acceptable; (c) Bob performs unitary operations on his qubits
according to the classical information received from Alice. Each
qubit is subjected to one of the four possible transformations
mentioned above.

Building on the case for one qubit (and also for two and three
qubits, which were analytically solved yet are too cumbersome to
be detailed here), we were able to induce the channel efficiency for
the N-qubit teleportation protocol:
\be
\label{C-protocol-N}
C^{pro}_N = \frac{2}{2^N+1}\left( 1 + \sum_{i=1}^N
2^{i-1}\textrm{Perm}_i^N\right),
\ee
where $\textrm{Perm}_i^N$ is the sum of all permutations of the
product of $i$ variables out of all \{$\chi_r$\}, where
$r=1,\ldots,N$, and $\chi_r =c(n_r)c(m_r)/2$. For example, the
three qubit case gives $\textrm{Perm}_1^3 = \chi_1+\chi_2+\chi_3$,
$\textrm{Perm}_2^3 = \chi_1\chi_2+\chi_1\chi_3+\chi_2\chi_3$, and
$\textrm{Perm}_3^3 = \chi_1\chi_2\chi_3$. We can better understand
Eq.~\eqref{C-protocol-N} by analyzing specific qubits to be
teleported. The contributions from $\textrm{Perm}_1^N$ appear when
we try to teleport product states, without entanglement. When
entangled qubits are teleported, the terms $\textrm{Perm}_{i>2}^N$
appear.

For the PQT of N qubits we see that $P_{suc}$ and
thus the protocol efficiency are $C^{PQT}_N=P_{suc}=\prod_{i=1}^N
2n_i^2/(1+n_i^2)^2$. Note that $C^{PQT}_N$ decreases
rapidly for a large number of qubits. This is due to the fact that
only measurement results that project Alice's qubits on
combinations of states given by $|\Phi^{-}_{m_{i}}\rangle$ and
$|\Psi^{+}_{m_{j}}\rangle$ yield unity fidelity. All the other
possible measurement outcomes are considered unacceptable in this
protocol and are discarded (they do not give unity fidelity).

\begin{figure}[!htb]
\centering
\includegraphics[width=8.4cm]{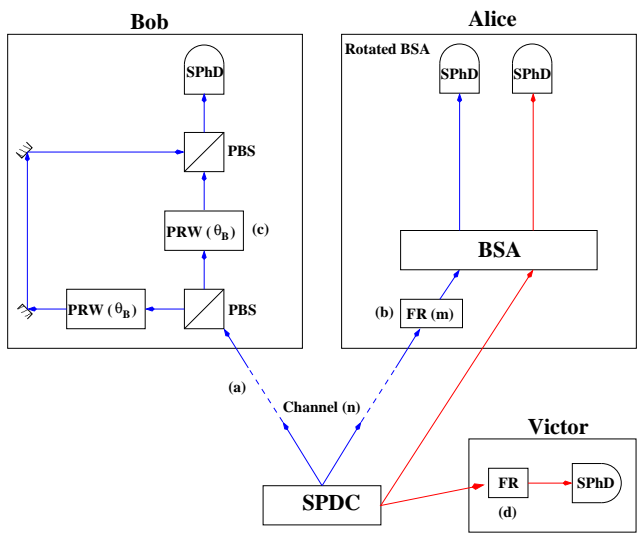}
\caption{(Color online) Scheme for experimental setup of the GTP.
%SPDC - Spontaneous Parametric Down-Conversion, FR - Faraday
%Rotator, SPhD - Single Photon Detectors, BSA - Bell State
%Analyzer, PBS - Polarization Beam Splitter, PRW - Phase
%Retardation (birefringent) Wave-plate.
(a) Channel entanglement given by parameter $n$;
(b) Measuring basis entanglement $m$;
(c) Bob's unitary operation degree of freedom $\theta_B$;
(d) Averaging capabilities by using different rotation angles.
See text for further details.}
\protect\label{Fig-Exp}
\end{figure}

A proposed experimental setup follows, using photon polarization
as the encoding medium. In this setup the horizontal $\ket{H}$ and
the vertical $\ket{V}$ polarizations of a photon encode $\ket{0}$
and $\ket{1}$, respectively. The features of the GTP are
incorporated into current standard teleportation experiments by
addition of simple linear photonic devices. The setup
(Fig.~\ref{Fig-Exp}) uses a Spontaneous Parametric Down-Conversion
(SPDC) \cite{kwi95} which emits two polarization-entangled photon
pairs. One pair is used as the teleporation channel, which can be
at an arbitrary length and purity thus facilitating different
channel entanglement $n$. The other pair is used for the
teleported qubit. By using Faraday Rotators (FRs) and
Single-Photon Detectors (SPhDs) Victor can measure one of the
photons along a given polarization axis, thus ascertaining the
other photon polarization and implement the averaging procedure.
One photon from the channel and the qubit to be teleported arrive
at Alice's, which has a complete Bell-State Analyzer (BSA). The
latter was recently shown to be implemented in several manners
\cite{wal05}. (Since it uses linear optics it is a probabilistic BSA 
\cite{VaiLut99}.) 
To transform it into a rotated-BSA (measures
arbitrary Bell-basis) a FR is inserted in the path of the qubit to
be teleported, introducing the free parameter $m$. The result of
the BSA is then transferred classically to Bob, who operates
unitarily on his qubit. This can be done by using Polarization
Beam Splitters (PBS) and Phase Retardation (birefringent)
Wave-plates (PRW). By using a PBS and two PRWs at the path of
Bob's photon we add another degree of freedom, namely
$\theta_{\Phi_\pm}=\theta_{\Psi_\pm}=\theta_B$. By measuring the
resulting photon polarization via a SPhD Bob ends the
teleportation protocol.

%%A simple application of the GTP in this experimental setup can
%%determine how the entanglement (or equivalently disentanglement)
%%of the channel affects the output fidelity. We can also find the
%%optimal parameters for maximal average teleportation fidelity. For
%%example, let us assume we select only the $\ket{\Psi^+_m}$ state
%%as an acceptable result and perform the
%%$e^{i\sigma_z\theta_B}\sigma_x$ unitary operation. By modifying
%%the FR at the rotated-BSA and the PRWs at Bob's, we can go over
%%all the range of $m$ and $\theta_B$. And averaging over many
%%teleported qubits results in a fidelity curve which depends on $m$
%%and $\theta_B$, with a maximum at
%%$|m|e^{i\theta_B}=|n|e^{i\theta_n}$. This maximum gives the
%%entanglement of the channel and, for a given set of acceptable
%%results, it also furnishes the optimal parameters from which the
%%highest average fidelity can be attained.

The generalized teleportation protocol (GTP) developed here shows
that the protocol efficiency depends solely on the channel and
measuring basis entanglement in a quite interesting way: {\em
interchanging the channel entanglement with the measurement basis
entanglement leaves the protocol efficiency unchanged} (See Eq.
(\ref{Cpro})). This result is also valid for the N qubit protocol.
The reason for this behavior
is unclear but we suspect that it is related to an unknown
information conservation law.
Furthermore, the protocol efficiency increases either if the channel
entanglement or the measurement basis entanglement are enhanced
and a deterministically unity fidelity protocol occurs only if
both quantities are maximum. We can understand this fact by noting
that quantum teleportation is a genuine quantum task relying on
entanglement. Therefore, as the availability of quantum resources
(entanglement) are increased, a better performance of the protocol
(higher output fidelity) is expected. On top of that, since the protocol
efficiency is given as a function of the
concurrences of the channels and of the measuring basis, we have
provided an operational physical interpretation for the
concurrence relating it to the efficiency of the teleportation
protocol.
Another result showed that for a dephased channel we can overcome
the dephasing easily by selecting the proper unitary
operation to counter it, implying that only the
entanglement of the channel is important to achieve a given output
fidelity and not which entangled state is used.
%We have also shown
%that decoherence control can be optimized by selecting the proper
%entanglement enhancement for each channel, according to
%Eq.~(\ref{C-protocol-N}).
%
Also, the generalization of the probabilistic quantum
teleportation protocol showed that for a given channel
entanglement, one can choose between different measurement basis
in order to {\em divide the channel efficiency between fidelity and
probability of success}. Finally, the extension of the protocol to
$N$ qubits gave new insights to the understanding of quantum
teleportation: the channel efficiency depends on the possible states
to be teleported. For an unentangled state, it only depends on the
individual channel concurrences, whereas for an entangled state it
depends on the product of the concurrences of the channels used in
the ``entanglement teleportation''.

%\begin{acknowledgments}
The authors thank G. Kurizki for his support and guidance.
G. R. thanks FAPESP and CNPq for funding.
%\end{acknowledgments}

\end{document}